%
%  SMALL SPHERES
%

\baselineskip 14pt plus 2pt

\font\llbf=cmbx10 scaled\magstep2
\font\lbf=cmbx10 scaled\magstep1

\def\ni{\noindent}

\def\ba{\bf a}
\def\bb{\bf b}

\def\uA{\underline A \,}      \def\bA{\bf A}
      \def\bB{\bf B}
      \def\bC{\bf C}
      \def\bD{\bf D}

\def\edth{\hskip 3pt {}^{\prime } \kern -6pt \partial }
\def\thorn{ { \rceil \kern -6pt \supset }}

%%%%%%%%%%%%%%%%%%%%%%%%%%%%%%%%%%%%%%%%%%%%%%%%%%%%%%%%%%%%%%%%%%%%%%%%%%%
%%%%%%%%%%%%%%%%%%%%%%%%%%%%%%%%%%%%%%%%%%%%%%%%%%%%%%%%%%%%%%%%%%%%%%%%%%%

\ni
{\llbf On certain quasi-local spin-angular momentum expressions for small 
spheres}\par
\bigskip
\bigskip
\ni
{\bf L\'aszl\'o B. Szabados}\par
\bigskip
\ni
Research Institute for Particle and Nuclear Physics \par
\ni
H-1525 Budapest 114, P.O.Box 49, Hungary \par
\ni
E-mail: lbszab@rmki.kfki.hu \par
\bigskip
\bigskip

\ni
The Ludvigsen--Vickers and two recently suggested quasi-local spin-angular 
momentum expressions, based on holomorphic and anti-holomorphic spinor 
fields, are calculated for small spheres of radius $r$ about a point $o$. 
It is shown that, apart from the sign in the case of anti-holomorphic 
spinors in non-vacuum, the leading terms of all these expressions coincide. 
In non-vacuum spacetimes this common leading term is of order $r^4$, and it 
is the product of the contraction of the energy-momentum tensor and an 
average of the approximate boost-rotation Killing vector that vanishes at 
$o$ and of the 3-volume of the ball of radius $r$. In vacuum spacetimes the 
leading term is of order $r^6$, and the factor of proportionality is the 
contraction of the Bel--Robinson tensor and an other average of the same 
approximate boost-rotation Killing vector.

\bigskip
\bigskip
\ni
{\lbf 1. Introduction}\par
\bigskip
\ni
In the classical theory of matter fields the energy-momentum and angular 
momentum {\it density} are described by the symmetric energy-momentum 
tensor $T^{ab}$. For Killing vectors $K^a$ the current $T^{ab}K_b$ is 
conserved, and it is interpreted as a component of the energy-momentum or 
angular momentum current density, depending on the nature of the Killing 
field. E.g. the translation- and rotation Killing vectors (or rather 
1-forms) of the Minkowski spacetime are given by $K^{\ba}_e:=\nabla_e
x^{\ba}$ and by $K^{\ba\bb}_e:=x^{\ba}K^{\bb}_e-x^{\bb}K^{\ba}_e$, where 
$x^{\ba}$ are the standard Descartes coordinates, and the corresponding 
currents are the four conserved energy-momentum and six conserved total 
(i.e. orbital- {\it and} spin-) angular momentum currents, respectively. 
By converting the name indices of the 
Killing 1-forms to spinor name indices the translations form a constant 
Hermitian matrix valued 1-form $K^{\bA\bA'}_e$, and the rotations can be 
decomposed as $K^{\bA\bA'\bB\bB'}_e=\varepsilon^{\bA'\bB'}K^{\bA\bB}_e+
\varepsilon^{\bA\bB}\bar K^{\bA'\bB'}_e$, the sum of the anti-self-dual 
and self-dual rotations, where e.g. the anti-self-dual rotations are given 
explicitly by $K^{\bA\bB}_e:=x^{(\bA}{}_{\bB'}K^{\bB)\bB'}_e$. If $\Sigma$ 
is any smooth spacelike hypersurface with fixed smooth 2-boundary $\$:=
\partial\Sigma$, $t^a$ is its future directed unit normal and ${\rm d}
\Sigma$ is the induced volume element, then $P^{\bA\bB'}_\$:=\int_\Sigma 
T^{ab}t_aK^{\bA\bB'}_b{\rm d}\Sigma$ and $J^{\bA\bB}_\$:=\int_\Sigma T^{ab}
t_aK^{\bA\bB}_b{\rm d}\Sigma$ depend only on the boundary \$ (and are 
independent of the actual $\Sigma$ defining the homology between \$ and 
zero), and hence may be interpreted as the quasi-local energy-momentum and 
angular momentum (with respect to the origin $o$ of the Descartes 
coordinate system) of the matter fields associated with \$, respectively. 
Let $r$ be a positive number and $\Sigma:=\{(t,x,y,z)\vert t=r\geq\sqrt{
x^2+y^2+z^2}\}$, i.e. the piece of the spacelike hyperplane $t=r$ that is 
bounded by its intersection with the future null cone ${\cal N}_o$ of the 
origin $o$, and let $\$_r:=\partial\Sigma$. Then the quasi-local 
energy-momentum and angular momentum can be computed for the sphere $\$_r$ 
of radius $r$ by taking the flux integrals of the conserved currents on 
$\Sigma$. But since $P^{\bA\bB'}_r$ and $J^{\bA\bB}_r$ depend only on $\$
_r$ and the origin $o$ is a regular point, they equal to the flux integral 
of the conserved currents on the null cone ${\cal N}_o$ between the vertex 
$o$ and $\$_r$. (For the technical details, e.g. the volume element on the 
null ${\cal N}_o$, see Section 3 below.) For sufficiently small $r$ the 
quasi-local energy-momentum is ${4\over3}\pi r^3(T^{ab}t_aK^{\bA\bB'}_b)
\vert_o$, the product of the 3-volume of $\Sigma$ and the value of $T^{ab}
t_aK^{\bA\bB'}_b$ at the origin. To compute the angular momentum too, it 
seems useful to introduce spinors. Let ${\cal E}^A_{\bA}:=\{O^A,I^A\}$ be 
the constant normalized spin frame field associated with the Descartes 
coordinates, ${\cal E}^{\bA}_A:=\{-I_A,O_A\}$ its dual basis, and 
$\varepsilon^A_{\uA}:=\{o^A,\iota^A\}$, $\varepsilon^{\uA}_A:=\{-\iota_A,
o_A\}$ the pair of standard dual spin frames on the null cone ${\cal N}_o$. 
If $\zeta:=\exp{\rm i}\phi\cot{\theta\over2}$, the complex stereographic 
coordinates (and hence $r,\zeta,\bar\zeta$ form a coordinate system on 
${\cal N}_o-\{o\}$), then $o^A(r,\zeta,\bar\zeta)={\rm i}{\root4\of{2}}(1+
\zeta\bar\zeta)^{-{1\over2}}(\zeta O^A+I^A)=:X^{\bA}{\cal E}^A_{\bA}$, and 
hence $X^{\bA}={\cal E}^{\bA}_Ao^A$. On the null cone the Hermitian matrix 
$x^{\bA\bB'}$ of the Descartes coordinates becomes $r$-times the dyadic 
product of the `spinor coordinates': $x^{\bA\bB'}=rX^{\bA}\bar X^{\bB'}$. 
Thus {\it on the null cone} the anti-self-dual rotation Killing field 
becomes 

$$
K^{\bA\bB}_e=ro^A{\cal E}^{(\bA}_A{\cal E}^{\bB)}_E\bar o_{E'}, \eqno(1.1)
$$
\ni
and, with accuracy $r^4$, the quasi-local anti-self-dual angular momentum 
for small spheres of radius $r$ is 

$$
J^{\bA\bB}_r={1\over4}r^4\bigl(T^{ab}{\cal E}^{(\bA}_C{\cal E}^{\bB)}_B
\bigr)\vert_o\oint_\$o_A\bar o_{A'}o^C\bar o_{B'}{\rm d}\$={4\pi\over3}
r^4T_{AA'BB'}t^{AA'}t^{B'E}\varepsilon^{BF}{\cal E}^{(\bA}_E{\cal E}^{\bB)}
_F\vert_o. \eqno(1.2)
$$
\ni
Here \$ is the unit sphere and ${\rm d}\$:= -2{\rm i}(1+\zeta\bar\zeta)
^{-2}{\rm d}\zeta{\rm d}\bar\zeta$, the 2-surface element on \$. Therefore 
the quasi-local angular momentum for small sphere of radius $r$ is ${1\over
4}r^3$ times the contraction of the energy-momentum tensor at $o$ and an 
average of the anti-self-dual rotation Killing vector that vanishes at $o$.
\par
      As a consequence of the diffeomorphism invariance of the geometric 
theories of gravity every vector field generates a conserved current (and 
its flux a conserved quantity), which can always be derived from a globally 
defined superpotential 2-form, independently of the homological structure 
of the spacetime [1]. In the case of energy-momentum in Einstein's theory this 
superpotential appears to be the Nester--Witten 2-form $u(\lambda,\bar\mu)
_{ab}$, associated with any pair of spinor fields $\lambda_A$ and $\mu_A$ 
[2,3], because of the following reasons. First, this 2-form defines a 
2-form on the spin frame bundle, and hence on the bundle of orthonormal 
frames over the spacetime manifold too, which 2-form extends uniquely to 
the bundle of linear frames $L(M)$. The superpotentials of the various 
classical energy-momentum pseudotensors (e.g. the Einstein, Bergmann, 
Landau-Lifshitz and the tetrad--M\o ller pseudotensors) are just the pull 
backs of various forms of this 2-form along various local cross sections 
of $L(M)$ [4,5]. In fact, the exterior derivative of the Nester--Witten 
form, known as Sparling's equation, looks like the Noether identity: 
$\nabla_{[a}u(\lambda,\bar\mu)_{bc]}=\Gamma(\lambda,\bar\mu)_{abc}-{1\over2}
\lambda_D\bar\mu_{D'}G^{de}{1\over3!}\varepsilon_{eabc}$, where $G_{ab}$ 
is the Einstein tensor and $\Gamma(\lambda,\bar\mu)_{abc}$, the so-called 
Sparling 3-form, is a quadratic expression of the derivatives of the spinor 
fields [6,7]. Second, both the ADM and Bondi--Sachs four-momenta can be 
written as the 2-surface integral of the Nester--Witten 2-form at spacelike 
and null infinity, where the spinor fields are chosen to be the constant or 
the asymptotic spinors there, respectively. These spinors may also be 
interpreted as the spinor constituents of the asymptotic translations at 
infinity. The simplest proof of the positivity of the ADM and Bondi--Sachs 
masses is probably that based on the use of the Nester--Witten 2-form 
[2,3,8-13]. Finally, the integral of the Nester--Witten 2-form for a 
spacelike topological 2-sphere \$ in the spacetime can be used to define 
energy-momentum, a manifest Lorentz covariant quantity (and not only energy, 
or energy and linear momentum separately), at the quasi-local level too. 
The only question is how to choose the two spinor fields $\lambda_A$ and 
$\mu_A$. Ludvigsen and Vickers [14] suggested a rule of transportation of 
the asymptotic spinors from infinity back to \$ if the spacetime is 
asymptoically flat at future null infinity and \$ can be connected with the 
future null infinity by a smooth null hypersurface. If the 2-surface \$ is 
a smooth spacelike cut of the future null cone of a point $o\in M$, then 
the two independent point spinors at $o$ can be transported to \$ by the 
same law of transportation [15], too. Dougan and Mason [16] suggested to 
choose anti-holomorphic or holomorphic spinor fields, which were shown to 
form two complex dimensional vector spaces in the generic case. Their 
constructions are genuine quasi-local in the sense that they can be applied 
in any spacetime for generic spacelike 2-surfaces which are homeomorphic to 
topological 2-spheres. The properties of these suggestions have 
been studied in a number of situations [14-20]. In particular, the 
Dougan--Mason energy-momentum is null if and only if the domain of 
dependence $D(\Sigma)$ of a spacelike hypersurface $\Sigma$ with smooth 
2-boundary \$ is a {\it pp}-wave geometry and the matter is pure radiation, 
and the energy-momentum is vanishing if and only if $D(\Sigma)$ is flat 
[18-20]. Furthermore, they have been calculated for small spheres of 
radius $r$ [15,17]. All the three expressions coincide in the leading 
order: in presence of matter they are ${4\over3}\pi r^3T^{ab}t_b$, in 
complete agreement with the expectation, whilst in vacuum (or if at least 
an open neighbourhood of the point $o$ is Ricci-flat) they are ${1\over
10G}r^5T^{abcd}t_bt_ct_d$, where $T_{abcd}$ is the Bel--Robinson tensor 
and $G$ is Newton's gravitational constant. The Dougan--Mason 
energy-momenta have also been calculated with accuracy $r^6$, and the two 
constructions deviate in this order. \par
      Because of the lack of any geometric meaning of the coordinates, the 
notion of angular momentum in general relativity is a more delicate problem, 
and it is argued e.g. in [21] that angular momentum should be connected with 
the internal Lorentz rotations of the theory, and hence should be analogous 
to the spin. In fact, general relativity has a Yang--Mills formulation 
(see e.g. [22,23]), in which Bramson derived the conserved Noether current 
corresponding to the Lorentz gauge symmetry and its superpotential 2-form 
[24] (and which has been rediscovered recently [25]). The Bramson spin 
superpotential 2-form, denoted here by $w(\lambda,\mu)_{ab}$, is well 
defined for any pair of spinor fields. Bramson used the asymptotic twistor 
equation to specify these spinor fields at future null infinity, and 
obtained an expression for the {\it global} spin-angular momentum by taking 
its integral on a spherical cut of null infinity. Thus to have a reasonable 
{\it quasi-local} spin-angular momentum associated with the 2-surface \$, 
the two spinor fields $\lambda_A$ and $\mu_A$ must be specifield there. 
For them it seems natural to use the spinor fields of the quasi-local 
energy-momentum. In fact, Ludvigsen and Vickers defined their quasi-local 
spin-angular momentum [14] as the integral of the Bramson superpotential 
with the spinor fields that they used in their energy-momentum expression. 
Recently it was suggested to use the 
anti-holomorphic or the holomorphic spinor fields in Bramson's superpotential, 
which yield genuine quasi-local spin-angular momentum expressions [20].
(Dougan and Mason suggested quasi-local angular momentum expressions, too, 
by using the spinor parts of the holmorphic or anti-holomorphic local 
twistors on \$ in the Nester--Witten 2-form [16]. In the present paper, 
however, their suggestion will not be investigated.) However we think that 
it is not enough to give a definition, but it should be clarified in various 
situations whether the new angular momentum expression really has the 
expected properties of the angular momentum, or, more generally, whether 
Bramson's superpotential serves an appropriate framework for finding the 
(quasi-local) measure of the angular momentum of gravity. In fact, in [20] 
the energy-momentum and spin-angular momentum expressions based on the 
anti-holomorphic unprimed spinors were calculated for finite axi-symmetric 
{\it pp}-wave Cauchy developments, and the energy-momentum was shown to be 
an eigenvector of the spin-angular momentum tensor. Therefore the null 
energy-momentum and the Pauli--Lubanski spin are proportional, a reasonable 
property that is shared by zero-rest-mass radiative matter fields in 
Minkowski spacetime. \par
      In the present paper the quasi-local Bramson spin-angular momenta 
are calculated for small spheres of radius $r$ for the Ludvigsen--Vickers 
and for the holomorphic and anti-holomorphic spinor fields. Since the 
angular momentum is the {\it momentum} of the energy-momentum, the leading 
orders are expected to be higher than those for the energy-momentum, 
namely $r^4$ for non-vacuum and $r^6$ for vacuum spacetimes. In fact, for 
non-vacuum spacetimes any reasonable {\it total} angular momentum expression 
can be expected to yield (1.2) as the leading term, simply because the 
energy-momentum tensor plays the role of the source for the gravity. The 
spin-angular momentum is a part of the total angular momentum, thus the 
same $r$-dependence is expected for the leading orders of the spin-angular 
momentum too. The prototypes of the small sphere calculations are those of 
Horowitz and Schmidt [26] for the Hawking energy [27] and of Kelly, Tod and 
Woodhouse [28] for the Penrose mass [29]. Recently Brown, Lau and York [30] 
calculated the Brown--York energy [31] for small spheres and increased the 
accuracy of some of the spin coefficients.\footnote{*}{For small spheres 
the Hawking and the Brown--York energies give 
the same result in the leading orders, namely ${4\pi\over3}r^3T_{ab}t^at^b$ 
in non-vacuum and ${2\over45G}r^5T_{abcd}t^at^bt^ct^d$ in vacuum. Thus, 
assuming that the weak energy condition is satisfied, the 
Ludvigsen--Vickers, the Dougan--Mason, the Hawking and the Brown--York 
energies are all {\it positive} for small spheres both in non-vacuum and 
vacuum spacetimes, which property can be interpreted as some form of the 
{\it local positivity} of the quasi-local energy.} 
Although the order of accuracy of these calculations is enough to calculate 
the quasi-local energy-momentum even in order $r^6$ (for the Dougan--Mason 
energy-momentum see [17]), the order of accuracy must be increased for the 
calculation of angular momenta. All the spin coefficients are needed with 
accuracy $r^3$. Thus first we review the main points of the framework of 
the calculations and present the spin coefficients and curvature components 
with $r^3$ accuracy. Section three is a quick review of the Nester--Witten 
and Bramson superpotentials, some of their properties and those forms of 
their integrals that we use. Since, as far as we know, the 
Ludvigsen--Vickers energy-momentum has not been calculated in vacuum in 
$r^6$ order, in Section 4 first we calculate this and compare with those of 
Dougan and Mason. Then the angular momenta will be calculated for the 
Ludvigsen--Vickers, the holomorphic and the anti-holomorphic spinors, both 
in non-vacuum and vacuum spacetimes. We will see that our expectation was, 
in fact, correct: In the non-vacuum case we recover (1.2) in all the three 
constructions, but with opposite sign for the anti-holomorphic spinors. In 
the vacuum case all the three construction give the same leading term. It 
is of order $r^6$ and the factor of proportionality is the contraction of 
the Bel--Robinson tensor and an other average of the approximate Killing 
vector (1.1). The results will be discussed and summarized in Section 5.\par
     Our conventions and notations are mostly those of [32]. In particular, 
we use the abstract index notations, and only the boldface and underlined 
indices take numerical values, e.g. ${\ba}=0,...,3$, ${\bA}=0,1$ and 
$\uA=0,1$. The spacetime signature is --2, the curvature- and Ricci tensors 
and the curvature scalar are defined by $-R^a{}_{bcd}X^bY^cZ^d:=\nabla_Y
(\nabla_ZX^a)-\nabla_Z(\nabla_YX^a)-\nabla_{[Y,Z]}X^a$, $R_{ab}:=R^c{}
_{acb}$ and $R:=R_{ab}g^{ab}$, respectively, and hence Einstein's equations 
take the form $G_{ab}:=R_{ab}-{1\over2}Rg_{ab}=-8\pi GT_{ab}$. On the other 
hand we use the GHP formalism in its original form [33], and we refer e.g. 
to equation (2.21) of [33] as (GHP2.21).

\bigskip
\bigskip
\ni
{\lbf 2. Small spheres}\par
\bigskip
\ni
First, mostly to fix the notations and ensure the coherence and readability 
of the paper, we summarize the geometric framework and philosophy of the 
small sphere calculations of [26] and especially of [28]. Since in this 
approximation the GHP equations have a hierarchical structure, they can be 
integrated with arbitrary high accuracy. In the second half of this section 
we integrate the GHP equations with the accuracy needed in Section 3, namely 
with $r^3$. \par
       Let $o\in M$ be an arbitrary point, $\{t^a,x^a,y^a,z^a\}$ be an 
orthonormal basis in $T_oM$ with $t^a$ future pointing, and parametrize the 
future celestial sphere at $o$ by $l^a(\theta,\phi):=t^a+x^a\sin\theta\cos
\phi+y^a\sin\theta\sin\phi+z^a\cos\theta$. They are precisely those null 
vectors at $o$ whose scalar product with $t^a$ is one. Let ${\cal N}_o:=
\partial I^+(o)$, the `future null cone' of $o$, let $l^a$ denote the 
tangent of its null geodesic generators coinciding with $l^a(\theta,\phi)$ 
at $o$ and let $r$ be the affine parameter along the integral curves of 
$l^a$, $l^a\nabla_ar=1$, and $r(o)=0$. Then in an open neighbourhood of $o$ 
the set ${\cal N}_o-\{o\}$ is a smooth null hypersurface and $\$_r:=\{p\in
{\cal N}_o\vert r(p)=r\}$ is a smooth spacelike 2-surface and homeomorphic 
to $S^2$. $\$_r$ is called a small sphere of radius $r$. $(r,\theta,\phi)$, 
or equivalently $(r,\zeta,\bar\zeta)$, forms a coordinate system on the 
smooth part of ${\cal N}_o$, where $\zeta:=\exp{\rm i}\phi\cot{\theta\over
2}$. Let us complete $l^a$ to a complex null tetrad $\{l^a,n^a,m^a,\bar m^a
\}$ such that $n^a$ is orthogonal to, and the complex null vectors $m^a$, 
$\bar m^a$ are tangent to the spheres of constant radius $\$_r$. If 
$\varepsilon^A_{\uA}:=\{o^A,\iota^A\}$ is the corresponding normalized 
spinor dyad on ${\cal N}_o$ such that $l^a\nabla_ao^B=0$, then, apart from 
constant transformations, the complex null tetrad (and hence the spinor 
dyad too) becomes fixed. The dual spin frame is $\varepsilon^{\uA}_A:=\{
-\iota_A,o_A\}$. The basis spinors $\{o^A(r,\zeta,\bar\zeta),\iota^A(r,
\zeta,\bar\zeta)\}$ (and the spinor components in this basis) have 
direction dependent limits at the vertex $o$. In fact, if ${\cal E}^A_{\bA}
:=\{O^A,I^A\}$ is the normalized spinor dyad at $o$ associated with the 
orthonormal vector basis $\{t^a,x^a,y^a,z^a\}$ (which we call Descartes 
spinors), then at $o$ 

$$
o^A(0,\zeta,\bar\zeta)={{\rm i}\root4\of{2}\over\sqrt{1+\zeta\bar\zeta}}
\Bigl(\zeta O^A+I^A\Bigr), \hskip 20pt
\iota^A(0,\zeta,\bar\zeta)={{\rm i}\over\root4\of{2}\sqrt{1+\zeta\bar
\zeta}}\Bigl(O^A-\bar\zeta I^A\Bigr). \eqno(2.1)
$$
\ni
In the coordinate system $(r,\zeta,\bar\zeta)$ the covariant directional 
derivative operators take the form $D:=l^a\nabla_a=({\partial\over\partial 
r})$, $\delta:=m^a\nabla_a=P({\partial\over\partial\bar\zeta})+Q({\partial
\over\partial\zeta})$ and $\bar\delta:=\bar m^a\nabla_a$, where $P$ and $Q$ 
are functions of the coordinates. The choice of the spinor dyad 
above yields the following relations on the GHP spin coefficients $\kappa=
\varepsilon=\rho-\bar\rho=\tau+\bar\beta'-\beta=\rho'-\bar\rho'=\tau'-
\beta'+\bar\beta=0$, and expressions for the GHP differential operators: 
$\thorn f=Df$, $\edth f=(\delta-(p-q)\beta-q\tau)f$ and ${\edth}'f=(\bar
\delta+(p-q)\bar\beta-p\bar\tau)f$, where $f$ has type $(p,q)$. In flat 
spacetime the relevant spin coefficients and the functions $P$ and $Q$ are 
$\kappa=\varepsilon=\tau=\tau'=\sigma=\sigma'=0$, $\rho=-{1\over r}$, 
$\rho'={1\over2r}$, $\beta=-{1\over2\sqrt{2}r}\zeta$, and $P={1\over\sqrt
{2}r}(1+\zeta\bar\zeta)$ and $Q=0$. Thus in flat spacetime $\delta$ reduces 
to ${1\over r}{}_0\delta$, where ${}_0\delta:={1\over\sqrt{2}}(1+\zeta\bar
\zeta)({\partial\over\partial\bar\zeta})$, and e.g. ${\edth}f$ reduces to 
${1\over r}{}_0{\edth}f$, where ${}_0{\edth}f:={}_0\delta f+{1\over2\sqrt2}
(p-q)\zeta f={1\over\sqrt2}(1+\zeta\bar\zeta)^{1-s}{\partial\over\partial
\bar\zeta}((1+\zeta\bar\zeta)^sf)$ with $s:={1\over2}(p-q)$, the spin weight 
of $f$. The Ricci identities (GHP2.21-24) and the primed version of 
(GHP2.25) and (GHP2.26), and the commutators (GHP2.31) and (GHP2.32) show 
that in general the deviation of the non-vanishing spin coefficients from 
these values is of order $O(r)$, e.g. $\rho=-{1\over r}+O(r)$, $\sigma=
O(r)$, and similarly $P={1\over\sqrt{2}r}(1+\zeta\bar\zeta)+O(r)$ and $Q=
O(r)$. Thus the flat space values of these quantities can be used as the 
initial data in finding the approximate solutions of accuracy $r^k$, 
$k\geq1$, of the GHP equations. The spin coefficients $\kappa'$ and 
$\varepsilon'$ and the operator ${\thorn}'$ do not play any role in the 
small sphere calculations; and, in fact, the geomtry of ${\cal N}_o$ does
not determine them. \par
      To find these solutions with accuracy $r^3$, first let us expand 
the spin coefficients and the functions $P$ and $Q$ as their flat space 
value plus polinomials up to third order in $r$, e.g. $\rho=-{1\over r}+
r\rho^{(1)}+r^2\rho^{(2)}+r^3\rho^{(3)}+O(r^4)$ and $P={1\over\sqrt{2}r}
(1+\zeta\bar\zeta)+rP^{(1)}+r^2P^{(2)}+r^3P^{(3)}+O(r^4)$, where the 
expansion coefficients are functions of $\zeta$ and $\bar\zeta$. Similarly, 
let us expand the curvature components retaining their first three 
non-trivial expansion coefficients, too, i.e. for example $\psi_0=\psi
^{(0)}_0+r\psi^{(1)}_0+r^2\psi^{(2)}_0+O(r^3)$. (At this point we should 
note that the different expansion coefficients have {\it different} $(p,q)$ 
types. For example $\psi^{(0)}_0$, $\psi^{(1)}_0$ and $\psi^{(2)}_0$ are of 
type (4,0), (5,1) and (6,2), respectively. Thus to save the extra care in 
using the GHP formalism we use the operators ${}_0\delta$ and ${}_0\bar
\delta$.) Then substituting the spin coefficients and curvature components 
into the GHP commutator (GHP2.31) (applied to the type (0,0) functions 
$\zeta$ and $\bar\zeta$) and the Ricci identities (GHP2.22-24), we get 

$$\eqalignno{
P&={1\over\sqrt2}\bigl(1+\zeta\bar\zeta\bigr)\Bigl\{{1\over r}+{1\over6}r
      \phi^{(0)}_{00}+{1\over12}r^2\phi^{(1)}_{00}+r^3\bigl[{7\over360}
      \bigl((\phi^{(0)}_{00})^2+\psi^{(0)}_0\bar\psi^{(0)}_{0'}\bigr)+{1
      \over20}\phi^{(2)}_{00}\bigr]\Bigr\}+O(r^4),&(2.2a)\cr
Q&={1\over\sqrt2}\bigl(1+\zeta\bar\zeta\bigr)\Bigl\{{1\over6}r\psi^{(0)}_0
      +{1\over12}r^2\psi^{(1)}_0+{1\over20}r^3\bigl({7\over9}\phi^{(0)}
      _{00}\psi^{(0)}_0+\psi^{(2)}_0\bigr)\Bigr\}+O(r^4),&(2.2b)\cr
\rho&=-{1\over r}+{1\over3}r\phi^{(0)}_{00}+{1\over4}r^2\phi^{(1)}_{00}+
      {1\over45}r^3\bigl( (\phi^{(0)}_{00})^2+9\phi^{(2)}_{00}+\psi^{(0)}_0
      \bar\psi^{(0)}_{0'}\bigr)+O(r^4),&(2.3a)\cr
\sigma&={1\over3}r\psi^{(0)}_0+{1\over4}r^2\psi^{(1)}_0+{1\over45}r^3\bigl(
      2\phi^{(0)}_{00}\psi^{(0)}_0+9\psi^{(2)}_0\bigr)+O(r^4),&(2.3b)\cr
\tau&={1\over3}r\bigl(\phi^{(0)}_{01}+\psi^{(0)}_1\bigr)+{1\over4}r^2
      \bigl(\phi^{(1)}_{01}+\psi^{(1)}_1\bigr)+\cr
    &+{1\over45}r^3\Bigl(2\bigl(\phi^{(0)}_{01}+\psi^{(0)}_1\bigr)\phi^{(0)}
      _{00}+2\bigl(\phi^{(0)}_{10}+\bar\psi^{(0)}_{1'}\bigr)\psi^{(0)}_0
      +9\bigl(\phi^{(2)}_{01}+\psi^{(2)}_1\bigr)\Bigr)+O(r^4).&(2.3c)\cr}
$$
\ni
Then from the commutator (GHP2.32) (applied either to $\zeta$ or to $\bar
\zeta$) the spin coefficient $\beta$ can be expressed by the expansion 
coefficients of the curvature {\it and} their ${}_0\delta$- and ${}_0\bar
\delta$-derivatives. To express the derivatives of the curvature components 
by the curvature components themselves, use the GHP Bianchi identity 
(GHP2.33). We get

$$\eqalign{
\beta=-&{1\over2\sqrt{2}r}\zeta+{1\over2}r\Bigl(\psi^{(0)}_1+{1\over6
     \sqrt{2}}\bigl(\bar\zeta\psi^{(0)}_0-\zeta\phi^{(0)}_{00}\bigr)\Bigr)
     +{1\over3}r^2\Bigl(\psi^{(1)}_1+{1\over8\sqrt{2}}\bigl(\bar\zeta\psi
     ^{(1)}_0-\zeta\phi^{(1)}_{00}\bigr)\Bigr)+\cr
   +&{1\over4}r^3\Bigl(\psi^{(2)}_1+{1\over10\sqrt{2}}\bigl(\bar\zeta\psi
     ^{(2)}_0-\zeta\phi^{(2)}_{00}\bigr)+{1\over18}\bigl(\psi^{(0)}_0\bar
     \psi^{(0)}_{1'}+4\psi^{(0)}_0\phi^{(0)}_{10}+3\phi^{(0)}_{00}\psi
     ^{(0)}_1\bigr)-\cr
    &-{7\over180\sqrt{2}}\bigl(\zeta\bigl((\phi^{(0)}_{00})^2+\bar\psi
     ^{(0)}_{0'}\psi^{(0)}_0\bigr)-2\bar\zeta\phi^{(0)}_{00}\psi^{(0)}_0
     \bigr)\Bigr)+O(r^4). \cr}\eqno(2.3d)
$$
\ni
Then the Ricci identity (GHP2.21) doesn't give anything 
new. To get the spin coefficients $\rho'$ and $\sigma'$ let us use the 
primed version of (GHP2.25) and (GHP2.26). They will be expressions of the 
expansion coefficients of the curvature and of their ${}_0\delta$- and 
${}_0\bar\delta$-derivatives. Some, but not all, of the derivatives of the 
curvature can be expressed by the curvature components themselves. First 
observe that the ${}_0\delta$- and ${}_0\bar\delta$-derivatives of the 
zeroth order curvature components, e.g. of $\psi^{(0)}_1$ and of $\phi^{(0)}
_{01}$, can be calculated from their definition using (2.1) above. 
(Geometrically, these formulae are not Bianchi identities, because they 
don't contain the derivatives of the curvature itself.) As a result 
$\rho'^{(1)}$ and $\sigma'^{(1)}$ become algebraic expressions of the 
curvature:

$$\eqalignno{
\rho'&={1\over2r}-{1\over6}r\Bigl(3\psi^{(0)}_2+3\bar\psi^{(0)}_{2'}+2\phi
       ^{(0)}_{11}-\phi^{(0)}_{00}+6\Lambda^{(0)}\Bigr)+O(r^2),&(2.3e') \cr
\sigma'&={1\over6}r\bigl(\bar\psi^{(0)}_{0'}-4\phi^{(0)}_{20}\bigr)+
       O(r^2).&(2.3f')\cr}
$$
\ni
To determine the higher order terms let us use the difference of the Bianchi 
identities (GHP2.34) and (GHP2.37), from which ${}_0\bar\delta\psi^{(1)}_1$ 
and ${}_0\bar\delta\psi^{(2)}_1$ can be expressed by the curvature 
components and by ${}_0\delta\phi^{(1)}_{10}$ and ${}_0\delta\phi^{(2)}
_{10}$, respectively. Then, using the definitions and (2.1) above, the 
remaining derivatives can be expressed by the 0-components of the 
(first and second) derivatives of the Weyl spinor and by the spinor 
components of certain irreducible parts of the derivatives of the Ricci 
spinor. Since in general these formulae are rather complicated and we do 
not need the general expressions, we concentrate only on the vacuum case. 
In vacuum we get 

$$\eqalignno{
\rho'=&{1\over2r}-{1\over2}r\bigl(\psi^{(0)}_2+\bar\psi^{(0)}_{2'}\bigr)-
      {1\over3}r^2\bigl(\psi^{(1)}_2+\bar\psi^{(1)}_{2'}\bigr)-\cr
    -&r^3\Bigl({1\over4}\bigl(\psi^{(2)}_2+\bar\psi^{(2)}_{2'}\bigr)-
      {1\over30}\psi^{(0)}_0\bar\psi^{(0)}_{0'}+{7\over90}\psi^{(0)}_1
      \bar\psi^{(0)}_{1'}\Bigr)+O(r^4),&(2.3e'') \cr
\sigma'=&{1\over6}r\bar\psi^{(0)}_{0'}+r^2\Bigl({1\over8}\bar\psi^{(1)}_{0'}
      -{1\over12}t^e\bigl(\nabla_e\bar\psi\bigr)_{0'}\Bigr)+\cr
    +&r^3\Bigl({1\over10}\bar\psi^{(2)}_{0'}-{1\over20}l^et^f\bigl(\nabla
      _{(e}\nabla_{f)}\bar\psi\bigr)_{0'}-{1\over10}\psi^{(0)}_2\bar\psi
      ^{(0)}_{0'}+{1\over15}\bar\psi^{(0)}_{2'}\bar\psi^{(0)}_{0'}-
    {11\over90}\bigl(\bar\psi^{(0)}_{1'}\bigr)^2\Bigr)+O(r^4). &(2.3f'')\cr}
$$
\ni
Here we used the notation $(\nabla_e\psi)_k$ and $(\nabla_{(e}\nabla_{f)}
\psi)_k$ for the $\nabla_e$- and $\nabla_{(e}\nabla_{f)}$-derivative of the 
Weyl spinor at $o$, respectively, contracted with $k$ $\iota^A$ and $(4-k)$ 
$o^A$ spinors. 
The remaining Ricci identity, namely the primed version of (GHP2.21), and 
the Bianchi identity (GHP2.34) do not restrict these components of the 
derivatives of the Weyl spinor further. \par
      By calculating the determinant of the induced 2-metric on 
$\$_r$, one gets the 2-area element on $\$_r$. It is ${\rm d}\$_r:=-{\rm i}
m_{[a}\bar m_{b]}=-{\rm i}(P\bar P-Q\bar Q)^{-1}{\rm d}\zeta\wedge{\rm d}
\bar\zeta$. Then $D({\rm d}\$_r)=-2\rho{\rm d}\$_r $, and hence if ${\rm d}
\$:=\lim_{r\rightarrow0}({1\over r^2}{\rm d}\$_r)=-2{\rm i}(1+\zeta\bar
\zeta)^{-2}{\rm d}\zeta\wedge{\rm d}\bar\zeta$, the area element of the unit 
metric sphere, then 

$$
{\rm d}\$_r=r^2\Bigl(1-{1\over3}r^2\phi_{00}^{(0)}-{1\over6}r^3\phi^{(1)}
_{00}-{1\over90}r^4\bigl[\psi^{(0)}_0\bar\psi^{(0)}_{0'}-4\bigl(\phi^{(0)}
_{00}\bigr)^2+9\phi^{(2)}_{00}\bigr]+O(r^5)\Bigr){\rm d}\$. \eqno(2.4)
$$
\ni
Thus in non-vacuum spacetimes ${\rm d}\$_r=r^2{\rm d}\$+O(r^4)$, whilst in 
vacuum ${\rm d}\$_r=r^2{\rm d}\$+O(r^6)$. \par
      Finally, by $t^{AA'}={1\over2}o^A\bar o^{A'}+\iota^A\bar\iota^{A'}$ 
the primed spin vectors at $o$ can be expressed by the unprimed ones: $\bar
o_{A'}=2t_{A'C}\iota^C$ and $\bar\iota^{A'}=t^{A'C}o_C$. Thus, combining 
these with Lemma (4.15.86) of [33], we get 

$$\eqalign{
\oint_{\$}o_{A_1}(\zeta,\bar\zeta)&...o_{A_k}(\zeta,\bar\zeta)\bar 
   o_{A'_1}(\zeta,\bar\zeta)...\bar o_{A'_l}(\zeta,\bar\zeta)\iota^{B_1}
   (\zeta,\bar\zeta)...\iota^{B_m}(\zeta,\bar\zeta)\bar\iota^{B'_1}(\zeta,
   \bar\zeta)...\bar\iota^{B'_n}(\zeta,\bar\zeta){\rm d}\$=\cr
&=2^lt_{A'_1C_1}...t_{A'_lC_l}t^{B'_1D_1}...t^{B'_nD_n}\oint_{\$}o_{A_1}...
   o_{A_k}o_{D_1}...o_{D_n}\iota^{B_1}...\iota^{B_m}\iota^{C_1}...\iota^{C_l}
   {\rm d}\$=\cr
&=\cases{{4\pi\over k+n+1}2^lt_{A'_1B_{m+1}}...t_{A'_lB_{m+l}}t^{B'_1A_{k+1}}
   ...t^{B'_nA_{k+n}}\delta^{(B_1}_{A_1}...\delta^{B_{m+l})}_{A_{k+n}}, &if 
   $k+n=l+m$;\cr
   0, &otherwise.\cr}\cr}\eqno(2.5)
$$
\ni
All the quasi-local energy-momentum and spin-angular momentum expressions 
for small spheres reduce to integrals of this type. \par
\bigskip
\bigskip

\ni
{\lbf 3. The Nester--Witten and Bramson superpotentials}\par
\bigskip
\ni
First recall that for any pair of spinor fields $\lambda_A$ and $\mu_A$ 
the Nester--Witten and Bramson 2-forms are defined by 

$$\eqalignno{
u(\lambda,\bar\mu)_{ab}:=&{{\rm i}\over2}\bigl(\bar\mu_{A'}\nabla_{BB'}
     \lambda_A-\bar\mu_{B'}\nabla_{AA'}\lambda_B\bigr),&(3.1)\cr
w(\lambda,\mu)_{ab}:=&-{\rm i}\lambda_{(A}\mu_{B)}\varepsilon_{A'B'}.
     &(3.2)\cr}
$$
\ni
If \$ is any orientable spacelike 2-surface in $M$ then their integral on 
\$ defines a Hermitian and a symmetric bilinear map, respectively, from 
$C^\infty(\$,{\bf S}_A)$, the (infinite dimensional) space of the smooth 
covariant spinor fields on \$, to the field of complex numbers. Thus if 
$\lambda^{\bA}_A$, ${\bA}=0,1$, is any pair of spinor fields on \$ then 
under the transformation of these spinor fields by {\it constant} $SL(2,
{\bf C})$ matrices the integrals 

$$
P^{\bA{\bB}'}:={1\over4\pi G}\oint_{\$}u(\lambda^{\bA},\bar\lambda
    ^{{\bB}'})_{ab}, \hskip 25pt 
J^{\bA\bB}:={1\over8\pi G}\oint_{\$}w(\lambda^{\bA},\lambda^{\bB})
    _{ab} \eqno(3.3,4)
$$
\ni
transform as a Hermitean and a symmetric spinor, respectively. Since the 
2-forms (3.1) and (3.2) 
are superpotentials for the energy-momentum and spin-angular momentum, for 
appropriately chosen spinor fields $\lambda^{\bA}_A$ these integrals are 
intended to define the quasi-local energy-momentum and the (anti-self-dual) 
spin-angular momentum, associated with the 2-surface \$, of the gravity plus 
matter system, respectively. Every two dimensional subspace of $C^\infty(\$,
{\bf S}_A)$ with some $SL(2,{\bf C})$ structure gives a potential 
definition for the quasi-local energy-momentum and spin-angular momentum. 
If one of the spinor fields in the arguments of the Nester--Witten 2-form, 
e.g. $\lambda^0_A$, is constant on \$ in the sense $m^e\nabla_e\lambda_A=0$ 
and $\bar m^e\nabla_e\lambda_A=0$ (whenever the spacetime geometry is 
considerably restricted [18-20]), then $u(\lambda^0,\bar\lambda^{{\bA}'})
_{ab}$ is exact, and hence the $00'$, $01'$ and $10'$ components of $P^{
{\bA}{\bA}'}$ are vanishing, and this $P^{\bA{\bA}'}$ is null with respect 
to the $SL(2,{\bf C})$ structure above. For this $P^{\bA{\bA}'}$ and the 
spin-angular momentum tensor $J^{\bA{\bA}'\bB{\bB}'}:=\varepsilon^{\bA\bB}
\bar J^{{\bA}'{\bB}'}+\varepsilon^{{\bA}'{\bB}'}J^{\bA\bB}$ we have $P_{\bA
{\bA}'}J^{\bA{\bA}'\bB{\bB}'}=(J^{01}+\bar J^{0'1'})P^{\bB{\bB}'}+(\delta
^{\bB}_0\delta^{{\bB}'}_{1'}J^{00}+\delta^{\bB}_1\delta^{{\bB}'}_{0'}\bar 
J^{0'0'})P^{11'}$. Thus the null $P^{\bA{\bA}'}$ is an eigenvector of 
$J^{\bA{\bA}'\bB{\bB}'}$ iff $J^{00}$ is vanishing; i.e. if and only if the 
pull back of $w(\lambda^0,\lambda^0)_{ab}$ to \$ is exact. However in 
general, without additional restrictions on $\lambda^1_A$ too, that is not 
exact. Mathematically, $J^{\bA\bB}$ is a measure on \$ of the 
non-integrability of the vector basis $E^a_{\bA{\bA}'}:=\lambda^A_{\bA}\bar
\lambda^{A'}_{{\bA}'}$. If both $\lambda^{\bA}_A$ are constant on \$, e.g. 
the restriction to \$ of the two covariantly constant spinor fields in 
Minkowski spacetime, then $w(\lambda^{\bA},\lambda^{\bB})_{ab}$ are exact 
and hence $J^{\bA\bB}$ is vanishing. Thus $J^{\bA{\bA}'\bB{\bB}'}$ is a 
measure how much the actual vector basis $E^a_{\bA{\bA}'}$ is `distorted' 
relative to the constant basis of the Minkowski spacetime.\par
       Next let us concentrate on the small spheres $\$_r$ of radius $r$ 
about $o$, and denote the corresponding integrals (3.3) and (3.4) by $P^{
{\bA}{\bB'}}_r$ and $J^{\bA\bB}_r$, respectively. Although these quantities 
can be calculated directly using the definitions (3.3) and (3.4), the formula 
(2.4) and the expansions of the spinor fields given explicitly in the next 
section, we can follow the philosophy of [26,15,17] by converting the 
2-surface integrals into integrals on the light cone too: Since for the 
Ludvigsen--Vickers, the anti-holomorphic and the holomorphic spinor fields 
$\lim_{r\rightarrow0}P^{\bA{\bB}'}_r=0$ and $\lim_{r\rightarrow0}J^{\bA\bB}
_r=0$, by the Stokes theorem these integrals can be calculated as the 
integrals of the exterior derivative of the superpotential 2-forms on 
${\cal N}^r_o$, the portion of the null cone ${\cal N}_o$ between the 
vertex and $\$_r$, too. To do the small sphere calculations in this way one 
must have a volume 3-form on the null hypersurface ${\cal N}_o$. It is 
chosen to be $\varepsilon_{abc}:=-{\rm i}3!n_{[a}m_b\bar m_{c]}$, because, 
for the naturally defined real tri-vector $\epsilon^{abc}:={\rm i}3!l^{[a}
m^b\bar m^{c]}$ on ${\cal N}_o-\{o\}$ and area 2-forms $\varepsilon_{ab}:=
-{\rm i}2m_{[a}\bar m_{b]}$ on the spacelike 2-surfaces $\$_r$ one has 
$\epsilon^{ijk}\varepsilon_{abc}=-\delta^{ijk}_{abc}$, $\varepsilon_{bc}=
l^a\varepsilon_{abc}$ and $\varepsilon_{abc}=3n_{[a}\varepsilon_{bc]}$, and 
hence the volume element 3-form on ${\cal N}_o$ is ${1\over3!}\varepsilon
_{abc}={1\over2}\varepsilon_{[ab}n_{c]}={\rm d}\$_r\wedge{\rm d}r$. Then the 
integral (3.3) takes the form  

$$\eqalign{
P^{\bA{\bB}'}_r={1\over4\pi G}\int_{{\cal N}^r_o}\nabla_{[a}&u(\lambda^{\bA},
  \bar\lambda^{{\bB}'})_{bc]}={1\over4\pi G}\int_0^r\oint_{\$_{r'}}\Bigl\{
  \bigl(D\lambda^{\bA}_A\bigr)\iota^A\bigl(\delta\bar\lambda^{{\bB}'}_{B'}
    \bigr)\bar o^{B'}+
  \bigl(D\bar\lambda^{{\bB}'}_{B'}\bigr)\bar\iota^{B'}\bigl(\bar\delta
    \lambda^{\bA}_A\bigr)o^A-\cr
-&\bigl(D\lambda^{\bA}_A\bigr)o^A\bigl(\bar\delta\bar\lambda^{{\bB}'}_{B'}
    \bigr)\bar\iota^{B'}-
  \bigl(D\bar\lambda^{{\bB}'}_{B'}\bigr)\bar o^{B'}\bigl(\delta\lambda^{\bA}
    _A\bigr)\iota^A+\cr
+&\bigl(\delta\lambda^{\bA}_A\bigr)o^A\bigl(\bar\delta\bar\lambda^{{\bB}'}
    _{B'}\bigr)\bar o^{B'}-
  \bigl(\bar\delta\lambda^{\bA}_A\bigr)o^A\bigl(\delta\bar\lambda^{{\bB}'}
    _{B'}\bigr)\bar o^{B'}-
  {1\over2}\lambda^{\bA}_E\bar\lambda^{{\bB}'}_{E'}G^{ef}o_F\bar o_{F'}
 \Bigr\}{\rm d}\$_{r'}{\rm d}r', \cr} \eqno(3.5)
$$
\ni
where we used the Sparling equation. Similarly, (3.4) can be written as 

$$\eqalign{
J^{\bA\bB}_r&={1\over8\pi G}\int_{{\cal N}_o^r}\nabla_{[a}w(\lambda^{\bA},
   \lambda^{\bB})_{bc]}=\cr
&={1\over8\pi G}\int_0^r\oint_{\$_{r'}}\Bigl\{D\bigl(\lambda^{\bA}_A
  \lambda^{\bB}_B\bigr)\bigl(o^A\iota^B+\iota^Ao^B\bigr)-2\bar\delta\bigl(
  \lambda^{\bA}_A\lambda^{\bB}_B\bigr)o^Ao^B\Bigr\}{\rm d}\$_{r'}{\rm d}r'.
  \cr} \eqno(3.6)
$$
\ni
For small enough $r$ these integrals can be expanded as a power series of 
$r$. To calculate these integrals with accuracy $r^k$, by (2.4) we need to 
calculate their integrand with accuracy $r^{(k-3)}$, and hence, because of 
the operator $D$, we need to calculate the spinor fields with accuracy 
$r^{(k-2)}$. However, as we will see explicitly, to compute (3.5) with 
accuracy $r^k$ it will be enough to calculate the spinor fields with accuracy 
$r^{(k-3)}$. This is due to the special nature of the Nester--Witten form. 
\par
\bigskip
\bigskip

\ni
{\lbf 4. Quasi-local spin-angular momenta for small spheres}\par
\bigskip
\ni
{\bf 4.1 The Ludvigsen--Vickers spinors}\par
\bigskip
\ni
First let us recall the definition of the Ludvigsen--Vickers spinors in the 
small sphere context [15]. They are the point spinors transported from $o$ 
to $\$_r$ along ${\cal N}_o$ by the propagation laws $(D\lambda_A)o^A=D
\lambda_0=0$, $(\bar\delta\lambda_A)o^A={\edth}'\lambda_0+\rho\lambda_1=0$, 
where the spinor components are defined by $\lambda_1o_A-\lambda_0\iota_A:=
\lambda_{\uA}\varepsilon^{\uA}_A:=\lambda_A$, and the two initial values for 
$\lambda_0$ at $o$ are the 0-components of the Descartes spinors ${\cal E}
^{\bA}_A=\{-I_A,O_A\}$: $\lambda^0_0(0):={\cal E}^0_Ao^A(0)={\rm i}
\root4\of{2}\zeta(1+\zeta\bar\zeta)^{-{1\over2}}$ and $\lambda^1
_0(0):={\cal E}^1_Ao^A(0)={\rm i}\root4\of{2}(1+\zeta\bar\zeta)^{-{1
\over2}}$. The first of the propagation laws implies that the spinor 
components $\lambda^{\bA}_0$ are independent of the affine parameter $r$, 
thus $\lambda^{\bA}_0$ on ${\cal N}_o$ are given explicitly by these 
expressions. (If a spinor has a name index, too, then its spinor components 
in the spin frame $\{\varepsilon^A_{\uA}\}$ will be written as subscripts 
and its name index as a superscript. Thus, for example, $\lambda^0_1$ is 
the 1-{\it component} of the {\it zeroth spinor}, $\lambda^0_1:=\lambda^0_A
\iota^A$, whilst $\lambda^1_0$ is the 0-{\it component} of the {\it first 
spinor}, $\lambda^1_0:=\lambda^1_Ao^A$.) Since the Ludvigsen--Vickers 
spinors are completely determined by the initial values, i.e. the Descartes 
spinors ${\cal E}^{\bA}_A$ at $o$, the metric $\varepsilon^{AB}$ on the 
space of point spinors at $o$ determines an $SL(2,{\bf C})$ structure on the 
space of the Ludvigsen--Vickers spinors on $\$_r$, too. It is this $SL(2,
{\bf C})$ structure that should be used to define the quasi-local mass as 
the length of $P^{\bA{\bB}'}$. In general the pointwise scalar product of 
the Ludvigsen--Vickers spinors, e.g. that for the basis spinors 
$\varepsilon^{AB}\lambda^{\bA}_A\lambda^{\bB}_B$, is {\it not} constant on 
$\$_r$. \par
     Following the general prescription of [26,28,15,17], let us expand the 
1-components as a power series of $r$. But since the operator $D$ in (3.5) 
and (3.6) reduces the power of $r$, to have $r^6$ accurate integrals, the 
spinor fields must be calculated with accuracy $r^4$: $\lambda^{\bA}_1=
\lambda^{\bA}_1{}^{(0)}+r\lambda^{\bA}_1{}^{(1)}+r^2\lambda^{\bA}_1{}^{(2)}
+r^3\lambda^{\bA}_1{}^{(3)}+r^4\lambda^{\bA}_1{}^{(4)}+O(r^5)$. Substituting 
this expansion into the second of the propagation laws and using (2.2) and 
(2.3) we get 

$$\eqalign{
\lambda^{\bA}_1{}^{(0)}&={}_0{\edth}'\lambda^{\bA}_0={}_0\bar\delta\lambda
       ^{\bA}_0-{1\over2\sqrt2}\bar\zeta\lambda^{\bA}_0, \cr
\lambda^{\bA}_1{}^{(1)}&=0, \cr
\lambda^{\bA}_1{}^{(2)}&={1\over2}\phi^{(0)}_{00}\lambda^{\bA}_1{}^{(0)}-
       {1\over3}\phi^{(0)}_{10}\lambda^{\bA}_0+{1\over6}\bar\psi^{(0)}
	_{1'}\lambda^{\bA}_0, \cr
\lambda^{\bA}_1{}^{(3)}&={1\over3}\phi^{(1)}_{00}\lambda^{\bA}_1{}^{(0)}-
       {1\over4}\phi^{(1)}_{10}\lambda^{\bA}_0+{1\over12}\bar\psi^{(1)}
	_{1'}\lambda^{\bA}_0, \cr
\lambda^{\bA}_1{}^{(4)}&={1\over4}\Bigl(\phi^{(2)}_{00}+{5\over6}\bigl(\phi
       ^{(0)}_{00}\bigr)^2+{1\over6}\psi^{(0)}_0\bar\psi^{(0)}_{0'}\Bigr)
	\lambda^{\bA}_1{}^{(0)}-\cr
     &-{1\over5}\Bigl(\phi^{(2)}_{10}-{1\over4}\bar\psi^{(2)}_{1'}-
       {1\over18}\bar\psi^{(0)}_{0'}\bigl(\phi^{(0)}_{01}-{11\over4}\psi
	^{(0)}_1\bigr)+{1\over9}\phi^{(0)}_{00}\bigl(7\phi^{(0)}_{10}-{19
	\over8}\bar\psi^{(0)}_{1'}\bigr)\Bigr)\lambda^{\bA}_0.\cr}
\eqno(4.1.1)
$$
\ni
Thus $\lambda^{\bA}_1{}^{(0)}$ are just the 1-components of the Descartes 
spinors: $\lambda^{\bA}_1{}^{(0)}={\cal E}^{\bA}_A\iota^A$. Therefore the 
spinor components $\lambda^{\bA}_0$ and $\lambda^{\bA}_1{}^{(0)}$ satisfy 
${}_0{\edth}\lambda^{\bA}_0=0$, ${}_0{\edth}\lambda^{\bA}_1{}^{(0)}+{1\over2}
\lambda^{\bA}_0=0$ and ${}_0{\edth}'\lambda^{\bA}_1{}^{(0)}=0$, too. The 
pointwise symplectic scalar product of the basis Ludvigsen--Vickers spinors 
is $\varepsilon^{AB}\lambda^{\bA}_A\lambda^{\bB}_B=\varepsilon^{\bA\bB}(1+{1
\over2}r^2\phi^{(0)}_{00}+{1\over3}r^3\phi^{(1)}_{00}+{1\over4}r^4[\phi^{(2)}
_{00}+{5\over6}(\phi^{(0)}_{00})^2+{1\over6}\psi^{(0)}_0\bar\psi^{(0)}_{0'}]
+O(r^5))$; i.e. the natural $SL(2,{\bf C})$ structure cannot be realized by 
the pointwise scalar product even in vacuum spacetimes. \par
     Before calculating the Ludvigsen--Vickers spin-angular momentum, for 
the sake of completeness let us calculate their energy-momentum with 
accuracy $r^6$ in vacuum spacetimes. As a consequence of the propagation 
laws, all the terms except the fifth in the integrand on the right hand 
side of (3.5) vanish. But, because of ${}_0{\edth}\lambda^{\bA}_0=0$, the 
integrand will be of order $r^2$, therefore to have $r^6$ accurate integral, 
by (2.4) it is enough to approximate ${\rm d}\$_r$ by $r^2{\rm d}\$ $. 
Finally, using (2.5), we get 

$$
P^{{\bA}{\bB}'}_r={1\over10G}r^5T^a{}_{bcd}t^bt^ct^d{\cal E}^{\bA}_A\bar
{\cal E}^{{\bB}'}_{A'}+{4\over45G}r^6t^e\bigl(\nabla_eT^a{}_{bcd}\bigr)t^b
t^ct^d{\cal E}^{\bA}_A\bar{\cal E}^{{\bB}'}_{A'}+O(r^7), \eqno(4.1.2)
$$
\ni
where $K^{\bA{\bB}'}_a:={\cal E}^{\bA}_A\bar{\cal E}^{{\bB}'}_{A'}\in T^*_oM$ 
may be interpreted as a translation at $o$ in the $\bA{\bB}'$ direction, and 
$T_{abcd}:=\psi_{ABCD}\bar\psi_{A'B'C'D'}$, the Bel--Robinson tensor at $o$. 
Thus the Ludvigsen--Vickers energy-momentum coincides with the Dougan--Mason 
energy-momentum based on the {\it holomorphic} unprimed spinors [17] even 
in sixth order. It is known [17], on the other hand, that at future null 
infinity it is the expression based on the {\it anti-holomorphic} rather 
than the holomorphic unprimed spinors that coincides with the Bondi--Sachs 
(and hence the Ludvigsen--Vickers) energy-momentum. Therefore the 
Ludvigsen--Vickers energy-momentum is interpolating between the 
anti-holomorphic Dougan--Mason energy-momentum for large spheres near the 
future null infinity and the holomorphic Dougan--Mason energy-momentum for 
small spheres. \par
     Next let us calculate the Ludvigsen--Vickers spin-angular momentum in 
non-vacuum spacetimes with accuracy $r^4$. Since by (4.1.1) the integrand 
on the right hand side of (3.6) is of order $r$, we may write ${\rm d}\$
_r=r^2{\rm d}\$ $, and, using Einstein's equations, a straightforward 
calculation yields 

$$
J^{{\bA}{\bB}}_r={1\over4}r^3T^{ab}\oint_{\$}l_aK^{\bA\bB}_b{\rm d}\$+
O(r^5)={4\pi\over3}r^4T_{AA'BB'}t^{AA'}t^{B'E}\varepsilon^{BF}{\cal E}
^{(\bA}_E{\cal E}^{\bB)}_F+O(r^5),\eqno(4.1.3)
$$
\ni
where $T^{ab}$ is the energy-momentum tensor at the vertex $o$, and the 
symmetric complex matrix valued 1-form $K^{\bA\bB}_e$ is given by (1.1). 
Therefore the Ludvigsen--Vickers spin-angular momentum for small 
spheres in non-vacuum spacetimes gives precisely the expected result (1.2): 
the leading order is $r^4$, and the factor of proportionality is the 
contraction of the energy-momentum tensor and the average of the null 
tangent of the light cone of $o$ and $K^{\bA\bB}_e$, a 1-form field that 
can be interpreted as the approximate anti-self-dual rotation Killing 
1-form that vanishes at $o$. We would like to stress that the 
energy-momentum tensor appears here in a rather non-trivial way, in 
contrast to the $r^3$ order calculations of the quasi-local 
energy-momentum, where the energy-momentum tensor was present explicitly in 
the exact expressions just because of the Sparling equation. 
The integrand of (3.6) is still of order $r$ in vacuum, thus to 
calculate the Ludvigsen--Vickers spin-angular momentum in vacuum with 
accuracy $r^6$ it is still enough to write ${\rm d}\$_r=r^2{\rm d}\$ $ 
by (2.4). But then the integrand should be calculated with accuracy $r^3$. 
A direct calculation yields 

$$\eqalign{
J^{\bA\bB}_r=&{1\over144\pi G}r^5T^{abcd}\oint_{\$}l_al_bl_cK^{\bA\bB}_d
     {\rm d}\$+O(r^7)=\cr
=&{4\over45G}r^6T_{AA'BB'CC'DD'}t^{AA'}t^{BB'}t^{CC'}t^{D'E}\varepsilon
     ^{DF}{\cal E}^{(\bA}_E{\cal E}^{\bB)}_F+O(r^7). \cr}\eqno(4.1.4)
$$
\ni
Therefore in vacuum the leading term is of order $r^6$, and the structure 
of this expression is the same that of (4.1.3) with the Bel--Robinson tensor 
of the gravitational `field', instead of the energy-momentum tensor of the 
matter fields. \par
\bigskip

\ni
{\bf 4.2 The holomorphic spinors}\par
\bigskip
\ni
Recall that a spinor field $\lambda_A$ on a spacelike 2-surface \$ is called 
holomorphic if $\bar m^b\nabla_b\lambda_A=0$, which in the GHP formalism is 
equivalent to ${\edth}'\lambda_1+\sigma'\lambda_0=0$ and ${\edth}'\lambda_0
+\rho\lambda_1=0$. If \$ is homeomorphic to $S^2$ then there are at least 
two, and for metric spheres there are precisely two linearly independent 
holomorphic spinor fields $\lambda^{\bA}_A$, ${\bA}=0,1$, for which 
$\varepsilon^{AB}\lambda^{\bA}_A\lambda^{\bB}_B$ is always constant on \$. 
Apart from certain exceptional 2-surfaces (e.g. future marginally trapped 
surfaces, i.e. for which the GHP spin coefficient $\rho$ is zero) the two 
spinor fields can be chosen to be normalized to the Levi-Civita symbol: 
$\varepsilon^{AB}\lambda^{\bA}_A\lambda^{\bB}_B=\varepsilon^{\bA\bB}$. For 
small perturbations of metric spheres, e.g. actually for small spheres, the 
number of the holomorphic spinor fields is still two, and they span the spin 
space at each point of \$ (see e.g. [19]). Therefore the pointwise 
symplectic scalar product defines a natural $SL(2,{\bf C})$ structure on 
the space of holomorphic spinor fields. \par
       Following the general prescription of the small sphere calculations 
let us expand the components of the holomorphic spinor fields: $\lambda
_{\uA}=\lambda_{\uA}{}^{(0)}+...+r^4\lambda_{\uA}{}^{(4)}+O(r^5)$. Then the 
condition of holomorphy above yields a hierarchical system of inhomogenious 
linear partial differential equations for these components: 

$$\eqalign{
{}_0{\edth}'\lambda_0{}^{(k)}-\lambda_1{}^{(k)}&=\sum_{l=0}^{k-2}\Bigl(
   C^{\uA}_l\lambda_{\uA}{}^{(l)}+D^{\uA}_l{\partial\lambda_{\uA}{}^{(l)}
   \over\partial\bar\zeta}+E^{\uA}_l{\partial\lambda_{\uA}{}^{(l)}\over
   \partial\zeta}\Bigr) \cr
{}_0{\edth}'\lambda_1{}^{(k)}&=\sum_{l=0}^{k-2}\Bigl(F^{\uA}_l\lambda_{\uA}
   {}^{(l)}+G^{\uA}_l{\partial\lambda_{\uA}{}^{(l)}\over\partial\bar\zeta}+
   H^{\uA}_l{\partial\lambda_{\uA}{}^{(l)}\over\partial\zeta}\Bigr), 
   \hskip 20pt k=0,...,4 \cr}\eqno(4.2.1)
$$
\ni
where $C^{\uA}_l$,...,$H^{\uA}_l$ are explicit expressions of the expansion 
coefficients of the GHP spin coefficients and the functions $P$ and $Q$; 
and they are vanishing in flat spacetime. Its general solution can be 
written as the sum of two spinor fields. The first, as we will see 
explicitly, is determined completely by the zeroth order holomorphic spinor 
fields, which turns out to be a linear combination of the Descartes spinors 
at $o$, and, in addition, its first order term is vanishing. Thus there are 
two such independent spinor fields, and they have the structure $\lambda_A=
a_{\bA}{\cal E}^{\bA}_A+r^2\lambda_A{}^{(2)}+r^3\lambda_A{}^{(3)}+...$, 
where $\lambda_A{}^{(2)},\lambda_A{}^{(3)},...$ are determined by $a_{\bA}
{\cal E}^{\bA}_A$ (see below). They are particular solutions of the 
inhomogenious system (4.2.1). The second is the general solution of the 
homogenious equations and turns out to be the sum of arbitrary complex 
combination of the Descartes spinors in each order, i.e. it has the form 
$rA^{(1)}_{\bA}{\cal E}^{\bA}_A+...+r^4A^{(4)}_{\bA}{\cal E}^{\bA}_A+
O(r^5)$. Hence the $r^k$ accurate homogenious solutions form a $2k$ 
dimensional vector space. However, as noted in [17] (and realized first in 
the case of 2-surface twistors by Kelly, Tod and Woodhouse [28]), these 
`spurious' solutions correspond to the lack of any canonical isomorphism 
between the space of holomorphic spinor fields on $\$_r$ and $\$_{r'}$ with 
different radii $r$ and $r'$; or, in other words, between the space of 
holomorphic spinor fields on $\$_r$ and the space of point spinors at $o$. 
Hence they should be `gauge solutions', and neither the integral of the 
Nester--Witten nor that of the Bramson 2-form should be sensitive to them. 
(This issue has not been discussed even for the Nester--Witten form.) 
To check this, let $\lambda_A$ be any spinor field whose zeroth and first 
order coefficients in its power series expansion are constant linear 
combinations of the Descartes spinors, $\lambda_A=a_{\bA}{\cal E}^{\bA}_A+
rb_{\bA}{\cal E}^{\bA}_A+r^2\lambda_A{}^{(2)}+...$, let $\gamma_A=r^k
A_{\bA}{\cal E}^{\bA}_A$, a $k$th order gauge solution, $k=1,2,...$, and 
determine the leading order in the integral of $u(\lambda,\bar\gamma)_{ab}$ 
and of $w(\lambda,\gamma)_{ab}$ on $\$_r$. ($\lambda_A$ has the structure 
of a particular solution for $b_{\bA}=0$ and of another gauge solution for 
$a_{\bA}=0$. However, we don't impose any further condition on $\lambda_A$.) 
Then the leading orders in these integrals are 

$$\eqalign{
\oint_{\$_r}u(\lambda,\bar\gamma)_{ab}&=
   \cases{O(r^{k+3}) & in non-vacuum, \cr   
          O(r^{k+5}) & in vacuum;     \cr}  \cr
\oint_{\$_r}w(\lambda,\gamma)_{ab}&=
   \cases{O(r^{k+4}) & in non-vacuum, \cr
          O(r^{k+6}) & in vacuum.     \cr}  \cr}\eqno(4.2.2)
$$
\ni
(To calculate these integrals explicitly we need to use the full (2.4).) 
Thus the expected leading terms, namely the $r^3$ and $r^4$ order terms in 
non-vacuum and the $r^5$ and $r^6$ terms in vacuum spacetimes for the 
integral of the Nester--Witten and Bramson 2-forms, respectively, are in 
fact not sensitive to the spurious solutions, i.e. they are really `gauge 
solutions'. However, without further conditions on the spinor fields 
$\lambda_A$, the next orders are already sensitive to them. But since it 
is only the first order gauge solutions that yield ambiguity in the next 
non-trivial orders by (4.2.2), it seems natural to impose the {\it gauge 
condition that the first order term in the spinor fields $\lambda_A$ be 
vanishing}; i.e. we exclude first order gauge solutions by hand. Note that 
although in general, without additional structures, it is meaningless to 
speak about ``how much homogenious solution is contained in the particular 
solution $\lambda_A$", because of the specific structure of the particular 
solutions (given below) this gauge condition is well defined. With this 
gauge condition the first two non-vanishing orders of the energy-momentum 
and spin-angular momentum will be independent of the remaining gauge 
solutions. The gauge independence of higher order terms cannot be ensured 
by similar, additional gauge conditions. \par
     The quotient of the space of the holomorphic spinor fields on $\$_r$ 
and that of the gauge solutions form in fact a two complex dimensional 
vector space. Up to the second order in arbitrary spacetime we can choose 
the following particular solutions: 

$$\eqalign{
\lambda^{\bA}_0&=o^A{\cal E}^{\bA}_A+r^2\Bigl(2\bigl(\phi^{(0)}_{01}+\phi
    ^{(0)}_{12}\bigr)\iota^A-\bigl(\phi^{(0)}_{11}+\phi^{(0)}_{22}-
    {1\over4}\phi^{(0)}_{00}\bigr)o^A\Bigr){\cal E}^{\bA}_A+O(r^3),\cr
\lambda^{\bA}_1&=\iota^A{\cal E}^{\bA}_A+r^2\Bigl(\bigl(\phi^{(0)}_{11}+
    \phi^{(0)}_{22}-{1\over4}\phi^{(0)}_{00}\bigr)\iota^A+{1\over3}\bigl(
    2\phi^{(0)}_{10}+{1\over2}\bar\psi^{(0)}_{1'}\bigr)o^A\Bigr){\cal E}
    ^{\bA}_A+O(r^3).\cr}
\eqno(4.2.3a)
$$
\ni
In the quotient space these form a basis which is normalized in the sense 
$\varepsilon^{AB}\lambda^{\bA}_A\lambda^{\bB}_B=\varepsilon^{\bA\bB}(1+
O(r^3))$. In vacuum these reduce to Dougan's solutions [17], and 
substituting them into (3.5) we recover the Dougan--Mason energy-momentum 
with $r^5$ accuracy. (Having imposed the gauge condition above, by (4.2.3) 
one could compute the Dougan--Mason energy-momentum in non-vacuum with 
$r^4$ accuracy. However we don't expect anything interesting in this order.) 
Substituting (4.2.3a) into (3.6) for the self-dual spin-angular momentum we 
get the expected result (1.2). \par
     To calculate the holomorphic Dougan--Mason energy-momentum and the 
holomorphic spin-angular momentum in $r^6$ order, we need to know the 
holomorphic spinor fields with $r^3$ and $r^4$ accuracy, respectively. In 
vacuum they are 

$$\eqalign{
\lambda^{\bA}_0=o^A{\cal E}^{\bA}_A&-{1\over18}r^4\Bigl(-{1\over8}\psi^{(0)}
   _0\bar\psi^{(0)}_{0'}+\psi^{(0)}_1\bar\psi^{(0)}_{1'}+3\psi^{(0)}_2\bar
   \psi^{(0)}_{2'}+4\psi^{(0)}_3\bar\psi^{(0)}_{3'}+2\psi^{(0)}_4\bar\psi
   ^{(0)}_{4'}\Bigr)o^A{\cal E}^{\bA}_A+\cr
  &+{1\over9}r^4\Bigl(\psi^{(0)}_0\bar\psi^{(0)}_{1'}+3\psi^{(0)}_1\bar\psi
   ^{(0)}_{2'}+4\psi^{(0)}_2\bar\psi^{(0)}_{3'}+2\psi^{(0)}_3\bar\psi^{(0)}
   _{4'}\Bigr)\iota^A{\cal E}^{\bA}_A+O(r^5),\cr
\lambda^{\bA}_1=\iota^A{\cal E}^{\bA}_A&+{1\over6}r^2\bar\psi^{(0)}_{1'}o^A
   {\cal E}^{\bA}_A+{1\over12}r^3\bar\psi^{(1)}_{1'}o^A{\cal E}^{\bA}_A+
   {1\over10}r^4\Bigl({1\over4}l^el^f\bigl(\nabla_e\nabla_f\bar\psi\bigr)
   _{1'}+{1\over3}\psi^{(0)}_1\bar\psi^{(0)}_{0'}\Bigr)o^A{\cal E}^{\bA}_A+\cr
  &+{1\over18}r^4\Bigl(-{1\over8}\psi^{(0)}_0\bar\psi^{(0)}_{0'}+\psi^{(0)}
   _1\bar\psi^{(0)}_{1'}+3\psi^{(0)}_2\bar\psi^{(0)}_{2'}+4\psi^{(0)}_3\bar
   \psi^{(0)}_{3'}+2\psi^{(0)}_4\bar\psi^{(0)}_{4'}\Bigr)\iota^A{\cal E}
   ^{\bA}_A+O(r^5).\cr}\eqno(4.2.3b)
$$
\ni
These are normalized with accuracy $r^4$: $\varepsilon^{AB}\lambda^{\bA}_A
\lambda^{\bB}_B=\varepsilon^{\bA\bB}(1+O(r^5))$. Substituting its $r^3$ 
accurate part into (3.5) we recover the result of [17], i.e. (4.1.2). 
Finally, substituting (4.2.3b) into (3.6), we obtain (4.1.4). Therefore 
the Ludvigsen--Vickers and the holomorphic expressions yield the same 
results even in $r^6$ order both for the energy-momentum and the spin-angular 
momentum. \par
\bigskip

\ni
{\bf 4.3 The anti-holomorphic spinors}\par
\bigskip
\ni
A spinor field $\lambda_A$ on \$ is called anti-holomorphic if $m^b\nabla_b
\lambda_A=0$, which in the GHP formalism is equivalent to ${\edth}\lambda_1
+\rho'\lambda_0=0$ and ${\edth}\lambda_0+\sigma\lambda_1=0$. The philosophy 
and the calculations are quite similar to those in the holomorphic case, 
thus we present only the results. Apart from the `spurious' solutions, for 
the independent, normalized anti-holomorphic spinor fields with accuracy 
$r^2$ in an arbitrary spacetime we can choose 

$$\eqalign{
\lambda^{\bA}_0&=o^A{\cal E}^{\bA}_A+r^2\Bigl(\psi^{(0)}_1\iota^A-\bigl(
    \psi^{(0)}_2-\Lambda^{(0)}\bigr)o^A\Bigr){\cal E}^{\bA}_A+O(r^3),\cr
\lambda^{\bA}_1&=\iota^A{\cal E}^{\bA}_A+r^2\Bigl(\bigl(\psi^{(0)}_2-
    \Lambda^{(0)}\bigr)\iota^A-\bigl(\psi^{(0)}_3-{1\over6}\bar\psi^{(0)}
    _{1'}-{1\over6}\phi^{(0)}_{10}\bigr)o^A\Bigr){\cal E}^{\bA}_A+O(r^3).\cr}
\eqno(4.3.1a)
$$
\ni
Substituting these into (3.6) for the anti-self-dual spin-angular momentum 
we get 

$$
J^{{\bA}{\bB}}_r=-{4\pi\over3}r^4T_{AA'BB'}t^{AA'}t^{B'E}\varepsilon^{BF}
{\cal E}^{(\bA}_E{\cal E}^{\bB)}_F+O(r^5),\eqno(4.3.2)
$$
\ni
i.e. that for the Ludvigsen--Vickers and the holomorphic spinors but with 
{\it opposite sign}. A pair of normalized holomorphic spinor fields in 
vacuum with accuracy $r^4$ is 

$$\eqalign{
\lambda^{\bA}_0&=o^A{\cal E}^{\bA}_A+r^2\Bigl(\psi^{(0)}_1\iota^A-\psi^{(0)}
    _2o^A\Bigr){\cal E}^{\bA}_A+{1\over3}r^3\Bigl(\bigl(\psi^{(1)}_1+2t^e
    (\nabla_e\psi)_1\bigr)\iota^A-\bigl(\psi^{(1)}_2+2t^e(\nabla_e\psi)_2
    \bigr)o^A\Bigr){\cal E}^{\bA}_A+\cr
  &+r^4\Bigl(-{1\over12}\bigl(l^el^f+2l^et^f+4t^et^f\bigr)\bigl(\nabla_{(e}
    \nabla_{f)}\psi\bigr)_2-{4\over9}\psi^{(0)}_1\psi^{(0)}_3+{1\over3}\bigl(
    \psi^{(0)}_2\bigr)^2\Bigr)o^A{\cal E}^{\bA}_A+\cr
  &+r^4\Bigl({1\over12}\bigl(l^el^f+2l^et^f+4t^et^f\bigr)\bigl(\nabla_{(e}
    \nabla_{f)}\psi\bigr)_1-{5\over36}\psi^{(0)}_0\psi^{(0)}_3+{1\over4}
    \psi^{(0)}_1\psi^{(0)}_2\Bigr)\iota^A{\cal E}^{\bA}_A+O(r^5),\cr
\lambda^{\bA}_1&=\iota^A{\cal E}^{\bA}_A+r^2\Bigl(\psi^{(0)}_2\iota^A-\bigl(
    \psi^{(0)}_3-{1\over6}\bar\psi^{(0)}_{1'}\bigr)o^A\Bigr){\cal E}^{\bA}_A
    +\cr
  &+{1\over3}r^3\Bigl(\bigl(\psi^{(1)}_2+2t^e(\nabla_e\psi)_2\bigr)\iota^A-
    \bigl(\psi^{(1)}_3+2t^e(\nabla_e\psi)_3\bigr)o^A+{1\over4}\bar\psi^{(1)}
    _{1'}o^A\Bigr){\cal E}^{\bA}_A+\cr
  &+r^4\Bigl(-{1\over12}\bigl(l^el^f+2l^et^f+4t^et^f\bigr)\bigl(\nabla_{(e}
    \nabla_{f)}\psi\bigr)_3+{1\over40}l^el^f\bigl(\nabla_{(e}\nabla_{f)}\bar
    \psi\bigr)_{1'}-\cr
  & \hskip 20pt -{1\over12}\psi^{(0)}_1\psi^{(0)}_4-{1\over36}\psi^{(0)}_2
    \psi^{(0)}_3+{7\over360}\psi^{(0)}_1\bar\psi^{(0)}_{0'}-{1\over6}\psi
    ^{(0)}_2\bar\psi^{(0)}_{1'}\Bigr)o^A{\cal E}^{\bA}_A+\cr
  &+r^4\Bigl({1\over12}\bigl(l^el^f+2l^et^f+4t^et^f\bigr)\bigl(\nabla_{(e}
    \nabla_{f)}\psi\bigr)_2-{5\over9}\psi^{(0)}_1\psi^{(0)}_3+{2\over3}\bigl(
    \psi^{(0)}_2\bigr)^2+{1\over6}\psi^{(0)}_1\bar\psi^{(0)}_{1'}\Bigr)
    \iota^A{\cal E}^{\bA}_A+O(r^5).\cr}
\eqno(4.3.1b)
$$
\ni
Substituting its $r^3$ accurate part into (3.5) we recover the result of 
[17], namely (4.1.2) with the numerical coefficient ${1\over9G}$ instead of 
${4\over45G}$ in the second term. (The $r^3$ accurate solution wasn't given 
in [17]].) Finally, substituting (4.3.1.b) into (3.6) we get (4.1.4). \par

\bigskip
\bigskip

\ni
{\lbf 5. Discussion and conclusions}\par
\bigskip
\ni
The integral $J^{\bA\bB}$ of the Bramson superpotential with the independent 
holomorphic or anti-holomorphic spinor fields on orientable 2-surfaces in 
arbitrary spacetimes and with the Ludvigsen--Vickers spinors in certain 
asymptotically flat spacetimes are well defined quasi-local observables of 
the gravity plus matter system. Their usefulness and interpretation as the 
spin-angular momentum, however, depends on their properties in specific 
situations, e.g. in the case of small spheres. To determine the general 
structure of the small sphere expression of such an integral one can follow 
the argument of Horowitz and Schmidt [26], applied originally to the Hawking 
energy: A simple dimension analysis shows that the number of derivatives of 
the metric in the coefficient of the $r^k$ term in the power series 
expansion of the quasi-local spin-angular momentum expressions is $(k-2)$. 
But the lowest order tensorial quantities as expansion coefficients appear 
for $k=4$ in non-vacuum and for $k=6$ in vacuum spacetimes, and, apart from 
numerical coefficients, these are the energy-momentum and the Bel-Robinson 
tensor, respectively, contracted with the unit vector $t^a$ and the 
generator of the physical quantity in question. The numerical coefficients 
may be different in the different constructions. \par
      The present calculations for the specific constructions confirm this 
`universal' structure of the small sphere expansions and the interpretation of
$J^{\bA{\bA}'\bB{\bB}'}$ as the gravitational angular momentum. Moreover, 
apart from the sign difference between the Ludvigsen--Vickers and holomorphic 
constructions on the one hand and the anti-holomorphic on the other in 
non-vacuum spacetimes, the results in the leading order are exactly the 
same in all the three constructions, and it is the expected result (1.2). 
The spin-angular momenta coincide in vacuum in the sixth order, and the 
factor of proportionality is just the Bel--Robinson tensor contracted with 
the unit vector $t^a$ and an average of the approximate boost-rotation 
Killing vector that vanishes at the origin. We stress that the approximate 
boost-rotation Killing vector {\it appears} at the end of the calculations, 
that is not put in the exact formulae by hand, in contrast to the 
translations in the small sphere calculations of the quasi-local 
energy-momentum. This result shows up the proper interpretation of 
the Bel--Robinson tensor as a quantity analogous to the energy-momentum 
tensor of the matter fields: it gives the main contribution to the 
gravitational self-energy-momentum, as has already been pointed out 
[26,28,15,17,30], {\it and} angular momentum at the quasi-local level. It 
might be worth noting that our previous result [20] that the {\it null} 
energy-momentum $P^{\bA{\bA}'}$ of an (axi-symmetric) domain of 
dependence $D(\Sigma)$ is an 
eigenvector of the angular momentum tensor $J^{\bA{\bA}'\bB{\bB}'}$ can 
easily be recovered in the small sphere approximation: If $\Sigma_r$ is 
any smooth hypersurface whose boundary is $\$_r$, the dominant energy 
condition is satisfied and $P^{\bA{\bA}'}_r$ is null, e.g. $P^{\bA{\bA}'}_r
=\delta^{\bA}_1\delta^{{\bA}'}_{1'}P^{11'}_r$, then $D(\Sigma_r)$ is 
known to be a {\it pp}-wave geometry 
and the matter is pure radiation [18,19]. Then by $o\in\overline{D(\Sigma
_r)}$ the energy-momentum and the Bel--Robinson tensor have the form $T_{AA'
BB'}=\vert\varphi\vert^2\lambda^0_A\bar\lambda^{0'}_{A'}\lambda^0_B\bar
\lambda^{0'}_{B'}$ and $T_{AA'BB'CC'DD'}=\vert\psi\vert^2\lambda^0_A\bar
\lambda^{0'}_{A'}\lambda^0_B\bar\lambda^{0'}_{B'}\lambda^0_C\bar\lambda
^{0'}_{C'}\lambda^0_D\bar\lambda^{0'}_{D'}$, respectively. Now it is easy 
to compute $P^{\bA{\bA}'}_r$ and, using (4.1.3) and (4.1.4), the angular 
momentum tensor $J^{\bA{\bA}'\bB{\bB}'}_r$ explicitly. Finally, one can 
check that $P^{\bA{\bA}'}_r$ is, in fact, an eigenvector of $J^{\bA{\bA}'
\bB{\bB}'}_r$. The quasi-local Pauli--Lubanski spin vector associated with 
the small sphere $\$_r$, defined by $S_{\bA{\bA}'}:={1\over2}\varepsilon
_{\bA{\bA}'\bB{\bB}'\bC{\bC}'\bD{\bD}'}P^{\bB{\bB}'}J^{\bC{\bC}'\bD
{\bD}'}$, is however vanishing in the expected leading oders, i.e. in $r^7$ 
and $r^{11}$, respectively. \par
    Since the angular momentum is not expected to have any definite sign 
(unlike the energy), the minus sign in 
(4.3.2) doesn't mean that the expression based on the anti-holomorphic 
spinors would be `non-physical'. The relative sign difference between the 
holomorphic and anti-holomorphic expressions shows only that, roughly 
speaking, the vector basis built from the holomorphic and anti-holomorphic 
spin frames are distorted relative to the constant basis by the curvature 
just in the opposite direction in $r^2$ order. \par
    Having imposed the gauge condition one can calculate the $r^5$ and 
$r^7$ order corrections. In order $r^5$ we don't expect anything 
interesting.  The $r^7$ accurate calculations for the spin-angular 
momentum would however be more interesting, because the holomorphic and 
anti-holomorphic constructions are expected to be different, but probably 
the Ludvigsen--Vickers and the holomorphic constructions are still coincide
in this order. But these calculations would be 
much more complicated because the spin coefficients and the spinor fields 
would have to be calculated with accuracy $r^4$ and $r^5$, respectively. 
\par
       In non-vacuum the result, apart from the sign in the anti-holomorphic 
case, is precisely the expected total quasi-local angular momentum of the 
matter fields. Therefore it is natural 
to think of the integral $J^{\bA\bB}$ with appropriately chosen spinor 
fields as the sum of the {\it total} angular momentum of the matter fields 
and the angular momentum of the gravity itself. But then the question arises 
whether that represents the total angular momentum of the gravitational 
field too, as claimed by some authors, or, in accordance with the 
expectation of Bramson, only its {\it spin part}; and in the latter case 
whether $J^{\bA\bB}$ should be completed by some quasi-local {\it orbital 
angular momentum} or not. Although the results of the present paper show 
that in the small sphere context the Bramson superpotential serves an 
appropriate framework for defining the gravitational angular momentum, in 
other situations it may yield wrong results. A general analysis shows that 
such an orbital term would have to be the integral of the momentum of the 
Nester--Witten superpotential [34], but it is not quite clear how this 
moment would have to be defined. To answer these questions one should 
consider other situations, e.g. large spheres near spacelike and/or null 
infinity, stationary axisymmetric systems etc, the subject of a future 
paper. \par
\bigskip
\bigskip

\ni
{\lbf Acknowledgement}\par
\bigskip
\ni
This work was partially supported by the Hungarian Scientific Research 
Fund grant OTKA T016246. 

\bigskip
\bigskip
\ni
{\lbf References}\par
\bigskip

\item{[1]} R.M. Wald, {\it On identically closed forms locally constructed 
           from a field}, J. Math. Phys. {\bf 31} 2378 (1990)
\item{[2]} E. Witten, {\it A new proof of the positive energy theorem},
           Commun. Math. Phys. {\bf 80} 381 (1981)
\item{[3]} J.M. Nester, {\it A new gravitational energy expression with
           a simple positivity proof}, Phys. Lett. {\bf 83 A} 241 (1981)
\item{[4]} J. Frauendiener, {\it Geometric description of energy-momentum
           pseudotensors}, Class. Quantum Grav. {\bf 6} L237 (1989)
\item{[5]} L.B. Szabados, {\it Canonical pseudotensors, Sparling's form 
           and Noether currents}, KFKI Report 1991-29/B, and its shortened 
	    version in Class. Quantum Grav. {\bf 9} 2521 (1992) 
\item{[6]} G.A.J. Sparling, {\it Twistors, spinors and the Einstein vacuum 
           equations}, University of Pittsburgh preprint, 1984
\item{[7]} M. Dubois-Violette, J. Madore, {\it Conservation laws and
           integrability conditions for gravitational and Yang--Mills field 
	    equations}, Commun. Math. Phys. {\bf 108} 213 (1987)
\item{[8]} W. Israel, J.M. Nester, {\it Positivity of the Bondi
           gravitational mass}, Phys. Lett {\bf 85 A} 259 (1981)
\item{[9]} M. Ludvigsen, J.A.G. Vickers, {\it The positivity of the Bondi
           mass}, J. Phys. A.: Math. Gen. {\bf 14} L389 (1981)
\item{[10]} M. Ludvigsen, J.A.G. Vickers, {\it A simple proof of the
            positivity of the Bondi mass}, J. Phys. A.: Math. Gen. {\bf 15}
            L67 (1982)
\item{[11]} O. Reula, {\it Existence theorem for solutions of Witten's
            equation and nonnegativity of total mass}, J. Math. Phys.
            {\bf 23} 810 (1982)
\item{[12]} G.T. Horowitz, K.P. Tod, {\it A relation between local and
            total energy in general relativity}, Commun. Math. Phys.
            {\bf 85} 429 (1982)
\item{[13]} O. Reula, K.P. Tod, {\it Positivity of the Bondi energy},
            J. Math. Phys. {\bf 25} 1004 (1984)
\item{[14]} M. Ludvigsen, J.A.G. Vickers, {\it Momentum, angular momentum
            and their quasi-local null surface extensions}, J. Phys. A:
            Math. Gen. {\bf 16} 1155 (1983)
\item{[15]} G. Bergqvist, M. Ludvigsen, {\it Quasi-local mass near a 
            point}, Class. Quantum Grav. {\bf 4} L29 (1987)
\item{[16]} A.J. Dougan, L.J. Mason, {\it Quasilocal mass constructions
            with positive energy}, Phys. Rev. Lett. {\bf 67} 2119 (1991)
\item{[17]} A.J. Dougan, {\it Quasi-local mass for spheres}, Class. Quantum
            Grav. {\bf 9} 2461 (1992)
\item{[18]} L.B. Szabados, {\it On the positivity of the quasi-local mass},
            Class. Quantum Grav. {\bf 10} 1899 (1993)
\item{[19]} L.B. Szabados, {\it Two dimensional Sen connections and
            quasi-local energy-momentum}, Class. Quantum Grav. {\bf 11}
            1847 (1994)
\item{[20]} L.B. Szabados, {\it Quasi-local energy-momentum and two-surface
            characterization of the pp-wave spacetimes}, Class. Quantum Grav.
            {\bf 13} 1661 (1996)
\item{[21]} P.G. Bergmann, R. Thomson, {\it Spin and angular momentum in 
            general relativity}, Phys. Rev. {\bf 89} 400 (1953)
\item{[22]} R. Utiyama, {\it Invariant theoretical interpretation of 
            interactions}, Phys. Rev. {\bf 101} 1597 (1956)
\item{[23]} T.W.B. Kibble, {\it Lorentz invariance and the gravitational 
            field}, J. Math. Phys. {\bf 2} 212 (1961)
\item{[24]} B.D. Bramson, {\it Relativistic angular momentum for 
            asymptotically flat Einstein--Maxwell manifolds}, Proc. Roy.
	     Soc. Lond. A. {\bf 341} 463 (1975)
\item{[25]} Y.-S. Duan, Sz.-S. Feng, {\it General covariant conservation law
            of angular momentum in general relativity}, Commun. Theor. Phys. 
            {\bf 25} 99 (1996)
\item{[26]} G.T. Horowitz, B.G. Schmidt, {\it Note on gravitational energy}, 
            Proc. Roy. Soc. Lond. A {\bf 381} 215 (1982)
\item{[27]} S.W. Hawking, {\it Gravitational radiation in an expanding
            universe}, J. Math. Phys. {\bf 9} 598 (1968)
\item{[28]} R.M. Kelly, K.P. Tod, N.M.J. Woodhouse, {\it Quasi-local mass 
            for small surfaces}, Class. Quantum Grav. {\bf 3} 1151 (1986)
\item{[29]} R. Penrose, {\it Quasi-local mass and angular momentum in
            general relativity}, Proc. Roy. Soc. Lond. {\bf A381} 53 (1982)
\item{[30]} J.D. Brown, S.R. Lau, J.W. York, {\it Canonical quasi-local 
            energy and small spheres}, gr-qc/9810003
\item{[31]} J.D. Brown, J.W. York, {\it Quasilocal energy and conserved 
            charges derived from the gravitational action}, Phys. Rev. {\bf 
	     D47} 1407 (1993)
\item{[32]} R. Penrose, W. Rindler, Spinors and Spacetime, vols. 1 and 2,
            Cambridge Univ. Press, Cambridge 1982 and 1986
\item{[33]} R. Geroch, A. Held, R. Penrose, {\it A space-time calculus 
            based on pairs of null directions}, J. Math. Phys. {\bf 14} 
	     874 (1973)
\item{[34]} L.B. Szabados, Talk given at the Erwin Schr\"odinger Institut, 
            Vienna, 1997 June, (unpublished)

\end